\documentclass[groupedaddress,
 reprint,
 amsmath,amssymb,
 nofootinbib,
 superscriptaddress,
 aps,
 prd,
]{revtex4-1}

\usepackage{graphicx}
\usepackage{dcolumn}
\usepackage{bm}
\usepackage{hyperref}
\usepackage[utf8]{inputenc}
\usepackage[capitalise]{cleveref}
\usepackage{aas_macros}
\usepackage[normalem]{ulem}
\usepackage{color}

\usepackage[makeroom]{cancel}
\usepackage{printlen}

\usepackage{mathtools}

\begin{document}

\title{Formation of inflaton halos after inflation}

\author{Benedikt Eggemeier}
\email{benedikt.eggemeier@phys.uni-goettingen.de}
\affiliation{
 Institut f\"ur Astrophysik, Georg-August-Universit\"at G\"ottingen, D-37077 G\"ottingen, Germany
}

\author{Jens C. Niemeyer}
\email{jens.niemeyer@phys.uni-goettingen.de}
\affiliation{
 Institut f\"ur Astrophysik, Georg-August-Universit\"at G\"ottingen, D-37077 G\"ottingen, Germany
}
\affiliation{Department of Physics, University of Auckland, Private Bag 92019, Auckland, New Zealand}

\author{Richard Easther}
\email{r.easther@auckland.ac.nz}
\affiliation{Department of Physics, University of Auckland, Private Bag 92019, Auckland, New Zealand}

\date{\today}

\begin{abstract}
The early Universe may have passed through an extended period of matter-dominated expansion following inflation and prior to the onset of radiation domination. Sub-horizon density perturbations grow gravitationally during such an epoch, collapsing into bound structures if it lasts long  enough. The strong analogy between this phase and structure formation in the present-day universe allows the use of N-body simulations and approximate methods for halo formation to model the fragmentation of the inflaton condensate into inflaton halos. For a simple model we find these halos have masses of up to $20\,\mathrm{kg}$ and radii of the order of $10^{-20}\,\mathrm{m}$,  roughly $10^{-24}$ seconds after the Big Bang. We find that the N-body halo mass function matches predictions of the mass-Peak Patch method and the Press-Schechter formalism within the expected range of scales. A long matter-dominated phase would imply that reheating and thermalization occurs in a universe with large variations in density, potentially modifying the dynamics of this process.  In addition, large overdensities can source gravitational waves and may lead to the formation of primordial black holes. 
\end{abstract}

\maketitle

\section{Introduction}
\label{sec:intro} 

The rapid expansion of the universe during inflation~\cite{Starobinsky1980,Guth1981,Linde1982,Linde1983} is followed by an epoch dominated by an oscillating inflaton field. In many cases the resulting condensate is rapidly fragmented by the resonant production of quanta, a process that depends on the detailed form of the inflaton potential and its couplings to other fields \cite{Shtanov:1994ce,Kofman1997,Lozanov:2016hid}. In the absence of resonance,   perturbations in the condensate laid down during inflation grow linearly with the scale factor after they re-enter the horizon and can collapse into bound structures prior to thermalization~\cite{Easther2010,Jedamzik:2010hq}.

It was recently demonstrated that the evolution and  gravitational collapse of the inflaton field during the post-inflationary era can be described by the non-relativistic Schrödinger-Poisson equations~\cite{Musoke2019}. This creates a strong analogy between the dynamics of the early universe and cosmological structure formation with fuzzy or axion-like dark matter; see for example~\cite{Schive:2014dra,Schive:2014hza,Veltmaat:2018dfz,Eggemeier2019PhR}. Building on this realisation, Ref.~\cite{Niemeyer2019} adapted the Press-Schechter approach to compute the mass function of the  gravitationally bound {\em inflaton halos\/} in the very early universe. These halos have macroscopic masses and microscopic dimensions: typical values are $\sim 0.01\,\mathrm{kg}$ with a virial radius of $\sim 10^{-22}\,\mathrm{m}$.  

The parallel with fuzzy dark matter structure formation suggests the existence of solitonic cores, or {\em inflaton stars} in the centers of inflaton halos with densities up to $10^6$ times larger than the average value, provided the reheating temperature is sufficiently low. Early bound structures can potentially  generate a stochastic gravitational wave background \cite{Jedamzik2010_grav} and runaway nonlinearities in the post-inflationary epoch can lead to the production of Primordial Black Holes (PBHs) whose evaporation could contribute to the necessary thermalization of the post-inflationary universe~\cite{GarciaBellido:1996qt,Anantua2009, Martin:2019nuw}.

We build upon the work presented in Refs.~\cite{Musoke2019,Niemeyer2019}, adapting a standard N-body solver to simulate the gravitational fragmentation of the inflaton field during the matter-dominated, post-inflationary epoch. In addition, we employ an approximate method to model halo formation based on the  mass-Peak Patch algorithm~\cite{Stein2018} to extend the dynamical range of the simulations.

\begin{figure*}
    \centering
    \includegraphics[width=\textwidth]{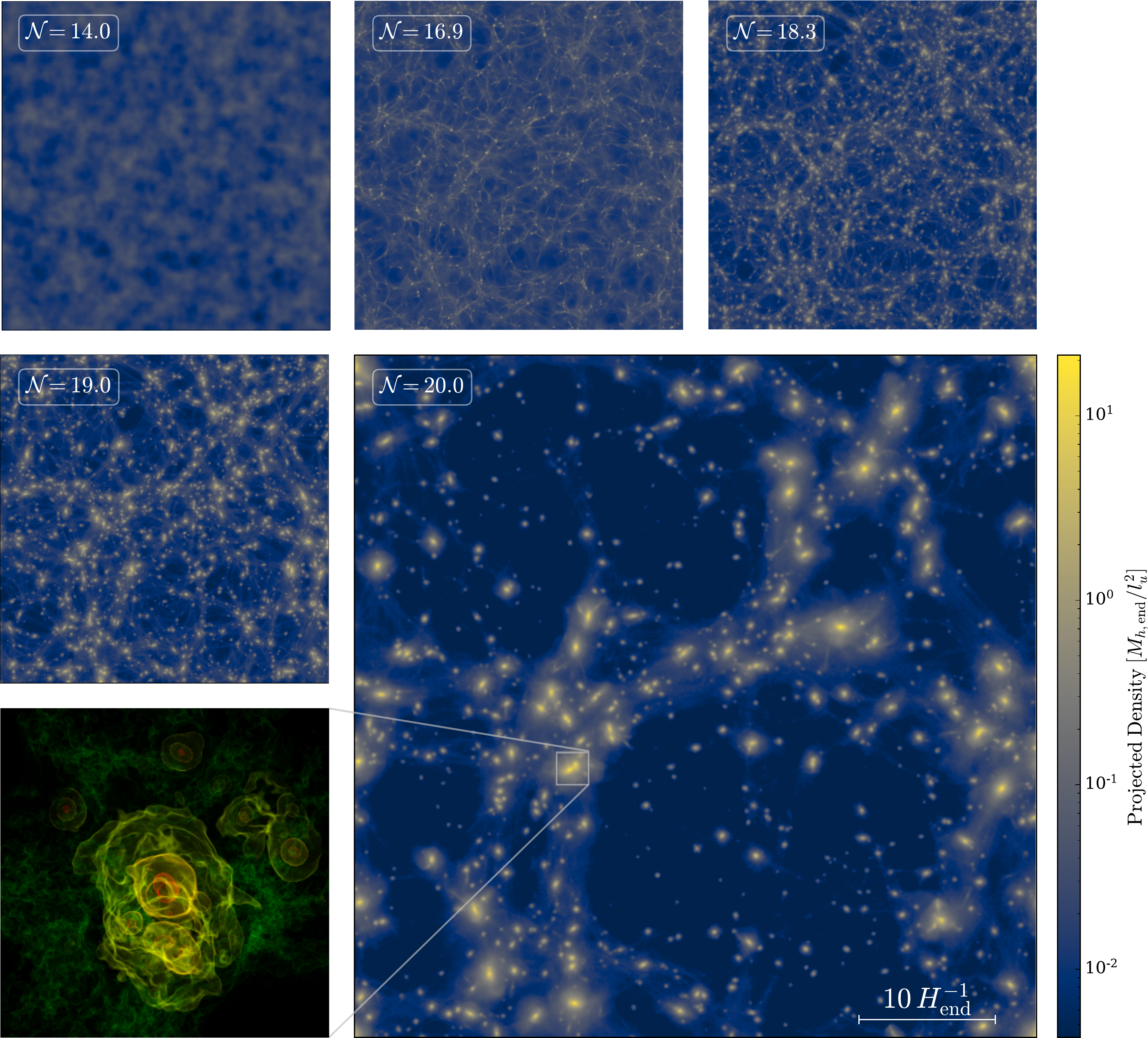}
    \caption{Projected inflaton density of the full simulation box from ${\cal{N}}=14$ $e$-folds (upper left) to ${\cal{N}}=20$ $e$-folds (lower right) after the end of inflation. Note that the color bar only applies to the final snapshot. The comoving length of the box is determined by the horizon size $H_\mathrm{end}^{-1}$ at the end of inflation. This corresponds to a physical size of $e^{20}\,H_\mathrm{end}^{-1}$ at the final snapshot of the simulations. A volume rendering of the largest inflaton halo with virial mass of $1.2\times 10^3\,M_{h,\mathrm{end}}\sim 20\,\mathrm{kg}$ is shown in the lower left panel. }
    \label{fig:dens_inflaton}
\end{figure*}

The N-body solver is initialised ${\cal{N}}=14$ $e$-folds after the end of inflation and run through to ${\cal{N}}=20$ $e$-folds, for a total growth factor of around 400. 
In contrast to the Schr\"{o}dinger-Poisson simulations of Ref.~\cite{Musoke2019}, it continues deep into the nonlinear phase, and we observe the formation and  subsequent growth and evolution of gravitationally bound structures (see \cref{fig:dens_inflaton}).  We  obtain the inflaton halo mass function (HMF) from the simulations, evaluating the density profiles of the halos, and analyze the density distribution of the inflaton field. 

While N-body solvers have an impressive dynamic range, they cannot capture the detailed dynamics of wavelike matter, i.e. the formation of solitonic cores and the surrounding incoherent granular density fluctuations. In particular, the initial power spectrum is suppressed at comoving scales below the post-inflationary horizon length \cite{Easther2010,Jedamzik:2010hq}, a situation in which N-body simulations generate spurious halos \cite{Wang2007,Lovell2014,Schneider2013,Schneider2015}. Fortunately, there are well-established procedures for filtering out these halos in dark matter simulations, which we can adapt to the early universe scenario considered here. 

During conventional structure formation the onset of dark energy domination puts an upper limit on the size of nonlinear objects. The effective matter-dominated phase in the early universe can last much longer than its present-day analogue, so the range of nonlinear scales may be much larger and the largest scales contained within our N-body simulation volume are becoming nonlinear as the calculation ends. However, we use \textsc{M3P}\footnote{Massively Parallel Peak Patches \cite{Behrens2020}}, a modified version of the mass-Peak Patch algorithm~\cite{Stein2018}, to cross-validate the N-body results  and explore the formation of collapsed structures at larger scales. 

This paper is structured as follows. We review single-field inflation and the early matter-dominated epoch that may follow it in \cref{sec:inflation}.  We describe the initial conditions and the simulation setup in \cref{sec:setup}; the results of the N-body and \textsc{M3P} calculations are presented in \cref{sec:results} and we combine these to yield an understanding of the inflaton HMF over a broad range of scales in the early universe. We conclude in \cref{sec:conclusions}.

\section{Inflation and early matter-dominated epoch}
\label{sec:inflation}
We consider single-field inflation in which the homogeneous inflaton scalar field $\varphi$ drives the expansion of the universe. In a flat Friedmann-Lemaitre-Robertson-Walker (FLRW) space-time, the evolution is described by the Friedmann equation
\begin{align}
    H^2 = \frac{1}{3M_\mathrm{Pl}^2}\left(\frac{1}{2}\dot\varphi^2 + V(\varphi)\right)\,,
    \label{eq:friedmann_eq}
\end{align}
where $H=\dot a/a$ is the Hubble parameter, $a$ is the scale factor, $M_\mathrm{Pl} = (8\pi G)^{-1/2}$ is the reduced Planck mass and $V(\varphi)$ denotes the effective potential of the scalar field $\varphi$. The inflaton obeys the Klein-Gordon equation
\begin{align}
    \Ddot{\varphi} + 3H\dot\varphi + V'(\varphi) = 0\,.
    \label{eq:kleingordon_eq}
\end{align}
As usual, a dot denotes a derivative with respect to cosmic time $t$ while a prime corresponds to a derivative with respect to $\varphi$. 
The universe expands roughly exponentially until the slow roll parameter $\varepsilon = -\dot H/H^2 = 1$, or  $(V'/V)^2/2 \approx 1$. We work with a quadratic potential,
\begin{align}
    V(\varphi) = \frac{1}{2}m^2\varphi^2\,.
    \label{eq:quadratic_potential}
\end{align}
Pure quadratic inflation is at odds with the data, but this can be regarded as the leading-order term near the minimum. We are assuming that the higher order terms in the potential will not support broad resonance~\cite{Lozanov:2017hjm}.

With a quadratic potential, \cref{eq:friedmann_eq,eq:kleingordon_eq} combine to deliver the well-known result 
\begin{align}
    \varphi(t) \sim \frac{M_\mathrm{Pl}}{m}\frac{\sin(mt)}{t}
\end{align}
in the post-inflationary epoch. Averaged over several oscillations, the scale factor grows as $a(t)\sim t^{2/3}$, thus $\varphi(t)\sim a^{-3/2}\sin(mt)$ and $H\sim a^{-3/2}$ \cite{Albrecht:1982mp}. Similarly, the energy density 
\begin{align}
    \rho_\varphi = \frac{1}{2}\dot\varphi^2 + \frac{1}{2}m^2\varphi^2
    \label{eq:energy_dens}
\end{align}
decreases as $\rho_\varphi\sim a^{-3}$, 
so the post-inflationary evolution can be treated as a matter-dominated universe on timescales larger than the frequency of the field oscillations. 

During this epoch density perturbations on sub-horizon scales initially grow  linearly, until they pass the threshold at which collapse becomes inevitable, leading to the formation of bound structures~\cite{Easther2010,Jedamzik:2010hq}. Modes that are only just outside the horizon at the end of inflation re-enter the horizon first, and are amplified the most. Hence the first nonlinear structures form on comoving scales slightly larger than scale of the horizon at the end of inflation.

The early matter-dominated epoch continues until the Hubble parameter becomes comparable to the effective decay rate $\Gamma$ of the inflaton. Once $H\simeq \Gamma$, reheating sets in and the inflaton decays into radiation. Since the decay rate is related to the reheating temperature~\cite{Kofman1997}
\begin{align}
    T_\mathrm{rh} \simeq 0.2(\Gamma M_\mathrm{Pl})^{1/2}\,,
    \label{eq:reheating_temp}
\end{align}
the energy scale at which inflation ends and the reheating temperature combine to determine the extent of the matter-dominated era. After the end of inflation the Hubble parameter evolves as $H\simeq H_\mathrm{end} (a/a_\mathrm{end})^{-3/2} = H_\mathrm{end}\exp{(-3{\cal{N}}/2)}$ where ${\cal{N}}$ denotes the number of $e$-folds after the end of inflation and the subscript ``end'' denotes a quantity evaluated at the end of inflation. Noting that $H\simeq \Gamma$, one obtains an estimate for the duration of the matter-dominated epoch,
\begin{align}
    {\cal{N}} \simeq \frac{2}{3}\ln\left(\frac{H_\mathrm{end}}{\Gamma}\right) \simeq \frac{2}{3}\ln\left(\frac{H_\mathrm{end}M_\mathrm{Pl}}{25T_\mathrm{rh}^2}\right)\,.
    \label{eq:Delta_N}
\end{align}

\section{Initial Conditions and Simulation Setup}
\label{sec:setup}

For definiteness  we set $m=6.35\times 10^{-6}\,M_\mathrm{Pl}$ and note that inflation ends\footnote{Similar assumptions are made in previous work~\cite{Easther2010,Niemeyer2019}.  Note that the pressure obeys $p=-\rho/3$ at the end of inflation,  or  $\rho_{\mathrm{end}} = 3V(\varphi_{\mathrm{end}})/2$ and  slow roll fails for quadratic inflation  when $\varphi \approx \sqrt{2} M_\mathrm{Pl}$. It is often generically assumed that a pivot scale $k_\ast=2\times 10^{-3}\,\mathrm{Mpc}^{-1}$ crosses the horizons ${\cal{N}}_\ast = 60$ $e$-folds before inflation ends but a long matter dominated phase reduces ${\cal{N}}_\ast$ \cite{Liddle2003,Adshead2008,Adshead2011} and $m$ is weakly dependent on the reheating scale, for a given normalisation. We ignore all these (small) corrections in what follows. } when $\varphi \approx M_{\mathrm{Pl}}$, so $H_\mathrm{end} \approx m/\sqrt{6}$. 
By definition, the Hubble horizon has radius $1/H$ and contains a mass
\begin{align}
    M_h =  \frac{4\pi M_\mathrm{Pl}^2 }{H}\,,
\end{align}
so $ M_{h,\mathrm{end}} = 0.021~\mathrm{kg}$.  
The physical size of the horizon at the end of inflation is $H_\mathrm{end}^{-1}$. In what follows,  physical quantities such as halo masses and length scales are given in units of $M_{h,\mathrm{end}}$ and $l_u \sim H_\mathrm{end}^{-1}$, respectively (see \cref{sec:unitsystem} for details).

\subsection{Initial Power Spectrum}

The power spectrum of density perturbations at the end of inflation was computed in Ref.~\cite{Easther2010} both numerically over a wide range of $k$ and analytically for super- and subhorizon scales. 
For scales that exit the horizon during inflation ($k < k_\mathrm{end}$) the slow-roll approximation can be employed and yields a weakly scale-dependent power spectrum. On  scales that never leave the horizon ($k > k_\mathrm{end}$) the dimensionless matter power spectrum obeys $\Delta^2 \sim k^{-5}$ (see \cref{sec:perturbation_ps} for details). The precise form of the power spectrum is not relevant for the nonlinear evolution of the density perturbations during the matter-dominated epoch~\cite{Musoke2019} and we interpolate between the sub- and superhorizon forms of the initial linear matter power spectrum~\cite{Niemeyer2019}.

\begin{figure}
    \centering
    \includegraphics[width=\columnwidth]{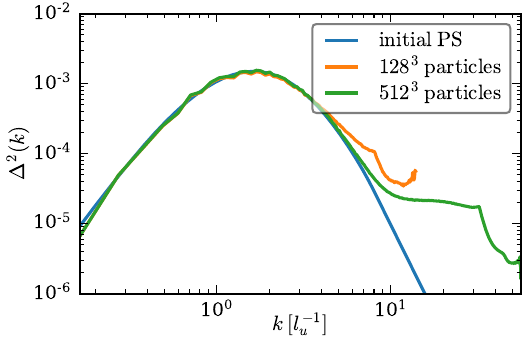}
    \caption{Initial input power spectrum ${\cal{N}}=14$ $e$-folds after the end of inflation in dimensionless units and computed power spectra from the initial density field for $128^3$ and $512^3$ particles, respectively. The increasing deviations for large $k$ from the input power spectrum arise due to discretisation effects (see~\cref{sec:ics_music} for details). }
    \label{fig:PS_initial}
\end{figure}

Density perturbations grow linearly with the scale factor, so we evolve the power spectrum forward from the end of inflation with the growth factor $D(a) \sim a$. Nonlinearities are expected to emerge after $\sim 17$ $e$-folds of growth \cite{Niemeyer2019}; we initialize the N-body solver 14 $e$-folds after inflation. The dimensionless power spectrum at this instant is shown in \cref{fig:PS_initial}.

\subsection{N-body Simulations}

In contrast to solving the full  Schr\"{o}dinger-Poisson dynamics~\cite{Musoke2019} N-body simulations can easily follow the evolution deep into the nonlinear phase. This provides an accurate understanding of halo formation and interactions, at the cost of obscuring small scale wavelike dynamics, including  solitonic cores in the inflaton halos and granular density fluctuations at the de Broglie scale.  In \cref{sec:results} we demonstrate the self-consistency of this approach by confirming that the de Broglie wavelength is significantly smaller than our spatial resolution.

We evolve the system through  six $e$-folds of growth, i.e. from ${\cal{N}}=14$ to ${\cal{N}}=20$. For a reheating temperature of $T_\mathrm{rh}\simeq 10^7\,\mathrm{GeV}$ the matter-dominated era following inflation lasts for ${\cal{N}} \simeq 24$ $e$-folds (cf.~\cref{eq:Delta_N}), which leaves plenty of scope for reheating and thermalization to take place before nucleosynthesis begins. 
The simulations  are performed using \textsc{Nyx}~\cite{Almgren2013}, with only gravitational interactions between the N-body particles.  The underlying dynamical system is identical to that which describes structure formation and evolution in a dark-matter-only universe but differs in parameter choices and power spectrum.   
The chosen length and time units are $l_u = 1.51\times 10^{-20}\,\mathrm{m}$ and $t_u = 7.23\times 10^{-24}\,\mathrm{s}$, with more details in \cref{sec:unitsystem}.

We generate the initial particle positions and velocities using \textsc{Music}~\cite{Hahn2011}, with the power spectrum shown in \cref{fig:PS_initial}. For comparison, the raw power spectra obtained from the initial density field with $128^3$ and $512^3$ N-body particles are also plotted. Except for resolution-dependent deviations which emerge at large $k$ the  power spectra coincide. For further details on the normalization of the input power spectrum in \textsc{Music} and a detailed discussion of the small scale form of the power spectra, see \cref{sec:ics_music}.
 
Two N-body simulations were performed, in comoving boxes with sides of length of $L=50\,l_u$ and $L=100\,l_u$ containing $512^3$ particles; the spatial resolution of the simulations is then $L/512$.  The tradeoffs between parameter choices are discussed in \cref{sec:resolution}.
Given that $\Omega_m=1$ throughout the simulation, the Hubble parameter at the end of the simulation $H_{20} = 6.49\,t_u^{-1}$, where the subscript 20 indicates the value of a parameter ${\cal{N}}=20$ $e$-folds after the end of inflation.  For a box size of $L=50\,l_u$ each initial Hubble region thus contains roughly $512^3/50^3 \approx 10^3$ N-body particles. Since the first halos form on comoving scales roughly equal to $H_\mathrm{end}^{-1}$ this is also the typical halo mass, and is much larger than the smallest resolvable halo in the simulation. 

Taking further advantage of the correspondence between dark matter and post-inflationary dynamics, we adapt the \textsc{Rockstar} halo finder~\cite{Behroozi2012} to locate the inflaton halos in the simulation outputs. The virial radius is then calculated from 
\begin{align}
    r_\mathrm{vir} = v_\mathrm{max} \, \left(\frac{4\pi}{3} G\rho_\mathrm{vir}\right)^{-1/2}\,,
\end{align}
where $\rho_\mathrm{vir} = \Delta_\mathrm{vir}\bar\rho$ with mean density $\bar\rho$, 
$\Delta_\mathrm{vir} = 18\pi^2$ for a matter-dominated background,
and $v_\mathrm{max}$ is the halo's maximum circular velocity. The virial mass of the halos is given by $M_\mathrm{vir} = 4\pi/3 \Delta_\mathrm{vir}\bar\rho r_\mathrm{vir}^3$.

\subsection{Mass-Peak Patch}

To validate the N-body simulations we apply \textsc{M3P} \cite{Behrens2020,Stein2018} 
to the initial density field. Rather than resolving the full nonlinear gravitational evolution, \textsc{M3P} identifies peaks in the linearly evolved density field corresponding to halos in an N-body simulation, generating large halo catalogues  using only a fraction of the CPU time and memory~\cite{Stein2018} required by an N-body code. \textsc{M3P} evolves the initial overdensity field with the linear growth factor $D(a) \sim a$ through to the final snapshot of the N-body simulation. Halo candidates in the density field are identified by smoothing with a top-hat filter on a hierarchy of filter scales, and based on the top-hat spherical collapse model an overdensity of $\delta_c = 1.686$ is the threshold at which a halo is selected. 
 
The Lagrangian radius, and hence halo mass, is determined by solving a set of homogeneous spherical collapse equations. To avoid double counting, halo candidates must be distinct, with no smaller collapsed objects contained within them and a hierarchical Lagrangian reduction algorithm is used to exclude overlapping patches from the halo catalogue. Finally, halo positions are computed via second order Lagrangian perturbation theory.
The filters must be chosen with care -- with too few filters viable halo candidates can be overlooked but having too many filters is computationally inefficient. Moreover, the set of filters has to span the mass range of the expected halos.

\section{Simulation Results}
\label{sec:results}

The N-body simulations show the formation and evolution of gravitationally bound inflaton halos. Visualizations of the full $L=50\,l_u$ simulation region at the beginning and end of the run, together with an enlargement of the largest halo, are shown in \cref{fig:dens_inflaton}. 
At ${\cal{N}}=20$ $e$-folds after the end of inflation, $\sim 60\%$ of the total mass is bound in inflaton halos with masses in the range  
$M_\mathrm{vir}\in [0.2,\,1.2\times 10^3]\,M_{h,\mathrm{end}}$ and corresponding virial radii $r_\mathrm{vir}\in [0.1,\, 2.0]\,l_u$. 

The spatial size of a solitonic core in the center of an inflaton halo is determined by its de Broglie wavelength $\lambda_\mathrm{dB} = 2\pi\hbar/(m v_\mathrm{vir})$, where $v_\mathrm{vir}$ is the virial velocity of the halo. The largest solitonic cores therefore exist in low-mass halos.  N-body simulations are unable to capture wavelike dynamics, even in principle. However, $\lambda_\mathrm{dB}\sim 10^{-4}\,l_u$ for a low-mass halo ${\cal{N}}=20$ $e$-folds after the end of inflation so even the most spatially extended solitons would be beneath the threshold for resolution by our simulations. On scales larger than $\lambda_\mathrm{dB}$, the Schrödinger-Poisson dynamics are governed by the Vlasov-Poisson equations justifying the use of N-body methods in this regime~\cite{Widrow1993,Uhlemann2014}.

\subsection{Halo Mass Function}
\label{sec:hmf}

\begin{figure*}
    \centering
    \includegraphics[width=\textwidth]{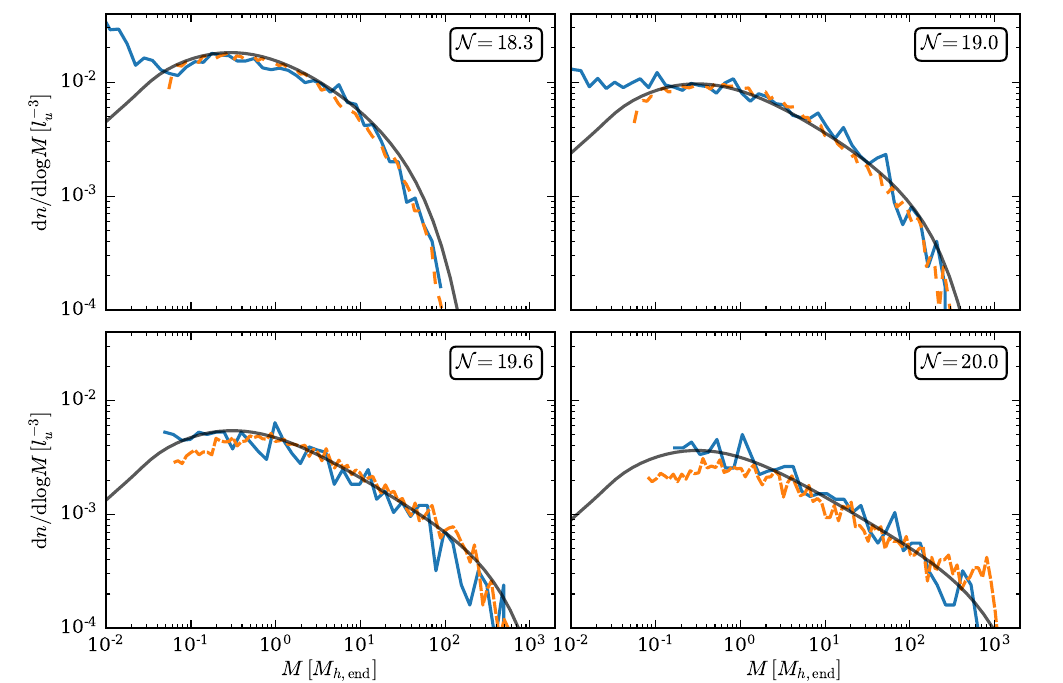}
    \caption{Evolution of the inflaton HMF. The blue solid lines represent the mass distribution of the halos identified in the $L=50\,l_u$ simulation box. Dashed orange lines display the HMF obtained via the \textsc{M3P} algorithm with box size of $L=100\,l_u$, while the black solid lines show the prediction from the Press-Schechter formalism with $\varepsilon=2.5$.}
    \label{fig:hmf}
\end{figure*}

N-body simulations with an initial power spectrum that has a well-resolved small-scale cutoff are known to produce spurious halos. These are found preferentially along filaments and can outnumber genuine physical halos below some mass scale. They are caused by artificial fragmentation of filaments and are common in warm dark matter (WDM) simulations~\cite{Wang2007,Lovell2014,Schneider2013,Schneider2015} where the free-streaming of particles induces a cutoff in the matter power spectrum. Via Eq.~(5) of Ref.~\cite{Lovell2014}, the scale below which spurious halos dominate is
\begin{align}
    M_\mathrm{lim} = 10.1\bar\rho d k_\mathrm{peak}^{-2}\,,
    \label{eq:mlim}
\end{align}
where $d$ is the effective spatial resolution 
and $k_\mathrm{peak}$ is the wave number at which the dimensionless initial power spectrum has its maximum (see \cref{fig:PS_initial}). 

With $L = 50\,l_u$ and $512^3$ particles, $d=9.8\times 10^{-2}\,l_u$ and we see from \cref{fig:PS_initial} that $k_\mathrm{peak} = 1.7\,l_u^{-1}$. Consequently, spurious halos dominate the HMF below $M_\mathrm{lim} \sim 0.1\,M_{h,\mathrm{end}}$. These small-scale halos are resolution-dependent and thus clearly unphysical~\cite{Wang2007} and must be filtered out of the halo catalogue. Algorithms that distinguish between artificial and genuine halos have been developed for WDM simulations\footnote{Instead of running a standard N-body simulation, it is also possible to trace dark matter sheets in phase space~\cite{Abel2012}, significantly suppressing the formation of spurious halos~\cite{Angulo2013}.} and we adapt a simplified but sufficient version for use here, as described in  \cref{sec:spurious_halos}. 

\begin{figure*}
    \centering
    \includegraphics[width=\textwidth]{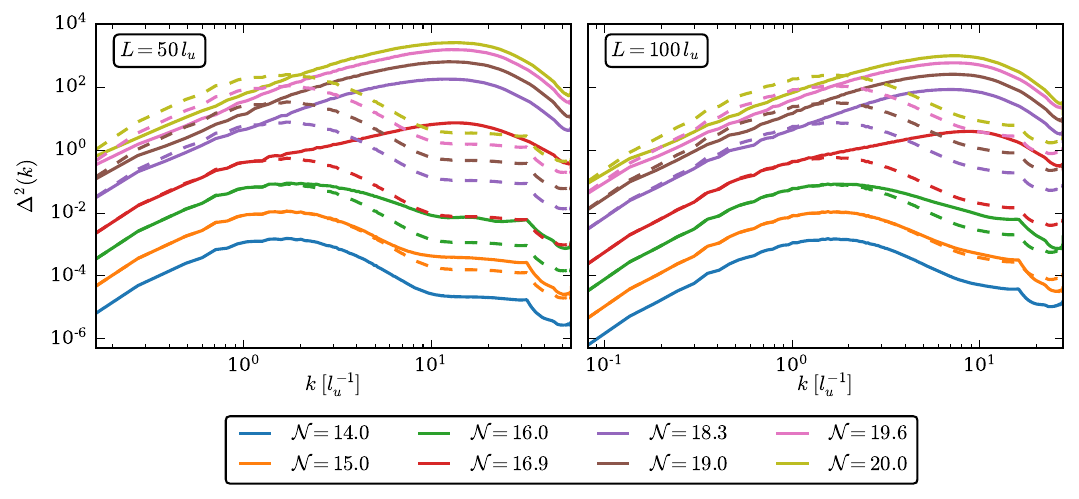}
    \caption{Dimensionless power spectrum for increasing ${\cal{N}}$ for a simulation box size of $L=50\,l_u$ (left) and $L=100\,l_u$ (right). The dashed lines display the \textsc{M3P} power spectrum, see \cref{eq:ps_z}.}
    \label{fig:PS}
\end{figure*}

The inflaton HMF is shown in \cref{fig:hmf} at four different times during the N-body simulation.  We compare the numerical HMFs with those computed using the \textsc{M3P} algorithm and Press-Schechter (PS) predictions~\cite{Press1974} with a sharp-$k$ filter \cite{Niemeyer2019,Schneider2013,Schneider2015}. 
This choice produces HMFs for power spectra with a cutoff at high wave numbers~\cite{Schneider2013,Schneider2015} and can be written as
\begin{align}
    \frac{\mathrm{d}n}{\mathrm{d}\ln M} = \frac{1}{6}\frac{\bar\rho}{M}\nu f(\nu) \frac{\Delta^2(1/R)}{\delta_c^2}\,,
    \label{eq:hmf_PS}
\end{align}
where
\begin{align}
    \Delta^2(k) = \frac{P(k)k^3}{2\pi^2}\,.
\end{align}
Following Ref.~\cite{Schneider2015}, we relate a mass $M$ to the filter scale $R$ via
\begin{align}
    M = \frac{4\pi}{3}\bar\rho(\varepsilon R)^3\,,
    \label{eq:mass_sharpk}
\end{align}
where the free parameter $\varepsilon$ has to be matched to simulations.   
We find that $\varepsilon=2.5$ yields Press-Schechter halo mass functions (PS-HMFs), displayed as black lines in \cref{fig:hmf}, that are in good agreement with the simulations at all times. In agreement with Ref.~\cite{Kulkarni2020}, we slightly rescaled the critical density $\delta_c$ in \cref{eq:hmf_PS} to  match the data (see \cref{sec:sharp_k} for details). 

Consistent with \cref{eq:mlim}, spurious halos dominate the N-body HMF for halos with masses lower than
$M_\mathrm{lim}\sim 0.1\,M_{h,\mathrm{end}}$, and account for the strong increase of the HMF for $M \lesssim M_\mathrm{lim}$, which becomes steeper for smaller ${\cal{N}}$.  
As discussed in \cref{sec:spurious_halos}, the removal of spurious halos is incomplete which explains the observed increase of the HMF at low masses.

However, the \textsc{M3P} results are unaffected by spurious low-mass halos and thus serve as a consistency check. Based on the mass range of the HMF from the N-body simulations we performed an \textsc{M3P} run with a total of 20 real space filters logarithmically spaced between $0.195\,l_u$ and $10\,l_u$. 
In order to adequately compare the \textsc{M3P} results to the N-body HMF we chose a larger box size of $L=100\,l_u$ for the \textsc{M3P} run (see \cref{sec:power_spectrum}).
The corresponding HMFs are shown as dotted lines in \cref{fig:hmf}. For $\mathcal{N}\leq 19$ they agree with the N-body HMFs over the entire mass range down to $M_\mathrm{lim}$. Notably, the low-mass end of the \textsc{M3P}-HMF is slightly underpopulated compared to both the N-body and the PS-HMF at $\mathcal{N}=19.6$, which becomes more pronounced at ${\cal{N}}=20$ when additionally more high-mass halos than expected are identified.
The reason why \textsc{M3P} reproduces the N-body and PS-HMF over the entire mass range at early times but cannot adequately do so at $\mathcal{N}\geq 19.6$ is that \textsc{M3P} identifies halos in the \emph{linearly} evolved density field and is hence not capable of including possible nonlinear contributions. This is further discussed in \cref{sec:power_spectrum}.

\begin{figure*}
    \centering
    \includegraphics[width=\textwidth]{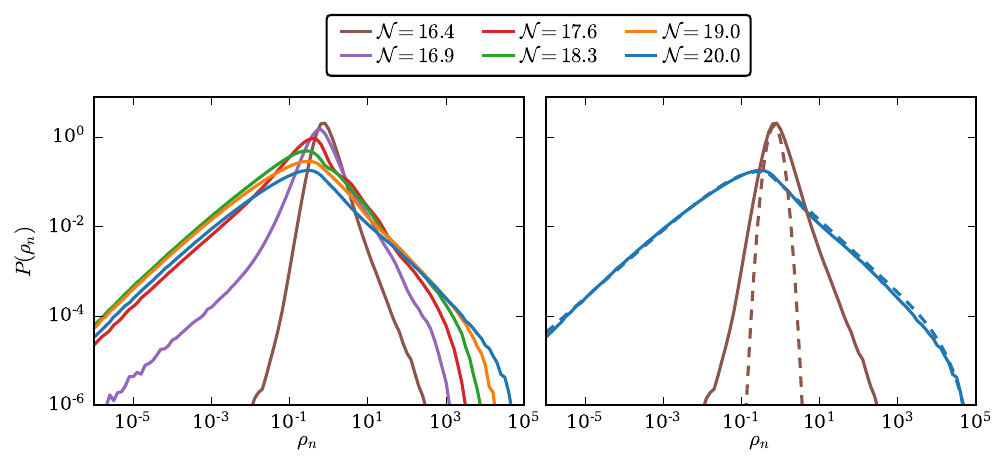}
    \caption{Left: one-point probability distribution function of the density field for increasing ${\cal{N}}$. Right: the same as on the left but with additional dashed lines that show a log-normal fit for $\mathcal{N}=16.4$ (see \cref{eq:fit_lognormal}) and a combined fit consisting of a power-law times double exponential fit for $\rho_n \leq 0.2$ (see \cref{eq:fit_left}) and a power-law times exponential fit for $\rho_n > 0.2$ for $\mathcal{N}=20.0$ (see \cref{eq:fit_right}), respectively.}
    \label{fig:dens_distribution}
\end{figure*}

\subsection{Power Spectrum}
\label{sec:power_spectrum}

Since we used different box sizes for the comparison of the inflaton HMFs in the previous section, we now analyze the power spectra with respect to different box sizes.
The evolving power spectrum of density fluctuations in dimensionless units is shown in \cref{fig:PS} for box sizes of $L=50\,l_u$ and $L=100\,l_u$.  Comparing the initial power spectrum of $L=50\,l_u$ at ${\cal{N}}=14$ with the one at ${\cal{N}}=16.9$ shows that the discretisation artefacts at large $k$, which are discussed in \cref{sec:ics_music}, are washed out at ${\cal{N}}=16.9$.
Unsurprisingly, we observe an overall increase in power with time, particularly for high-$k$ modes. 

The dotted lines in \cref{fig:PS} show the \textsc{M3P} power spectrum. It is related to the initial power spectrum via the linear growth factor $D(a)$:
\begin{align}
    \Delta^2(k,a) = \left(\frac{a}{a_\mathrm{init}}\right)^2\Delta^2(k,a_\mathrm{init})\,.
    \label{eq:ps_z}
\end{align}

As expected, the N-body and \textsc{M3P} power spectra coincide for small $k$, i.e. in the linear regime for lower values of  ${\cal{N}}$. 
However, at $\mathcal{N}\geq 19$ the \textsc{M3P} spectra differ significantly from the N-body spectra for $L=50\,l_u$ even at small $k$, indicating that all scales are now nonlinear. Thus, the linearly evolved \textsc{M3P} power spectrum for $L=50\,l_u$ at $\mathcal{N}\geq 19$ should not be used to obtain the corresponding HMF; an \textsc{M3P} run with a larger box size can resolve this issue though. 

As expected, the $L=100\,l_u$ \textsc{M3P} spectra shown in the right panel of \cref{fig:PS} agree with the N-body power spectra at small $k$ for a longer time, however slight deviations at large scales are observable at $\mathcal{N}\gtrsim 19.6$. This leads to inaccuracies in the \textsc{M3P}-HMF in a sense that compared to the N-body results and the PS-HMF more high-mass and less low-mass halos are predicted at $\mathcal{N}=20$, as can be seen in the lower right panel of \cref{fig:hmf}.

\subsection{Density Distribution}
We determine the density distribution of the matter field  (i.e. the one-point probability distribution function) by binning the normalized density $\rho_n =  \rho/\bar\rho = 1 + \delta$, where $\delta$ is the overdensity, using a logarithmic bin width  $\Delta \log(\rho_n) = 0.1$. 
The density distribution function $P(\rho_n)$  illustrates the relative frequency of overdensities. It is defined to be the normalized number of cells whose corresponding density value lies in a range by $\Delta \log(\rho_n)$, i.e. $P(\rho_n) = \Delta N_\mathrm{cell}/ \Delta \log(\rho_n) N_\mathrm{cell}^{-3}$ where $N_\mathrm{cell}=512$ is the grid size.   

The evolution of the density distribution in the $L=50\,l_u$ simulation is shown in the left panel of \cref{fig:dens_distribution}. It initially is a narrow distribution, reflecting the shape of the power spectrum, which widens as the nonlinear phase continues. The observed maximal overdensity increases, due to ongoing gravitational collapse, mergers, and accretion onto existing halos. As a consequence, the number of cells with a low mass density increases in order to supply the raw material for the growing overdensities. 
At ${\cal{N}}=20$ the densities range from roughly $10^{-6}$ to nearly $10^5$, with and the distribution peaks at $\rho_n \sim 0.2$. 

We now approximate the density distribution functions that we obtained from the numerical simulations. As a starting point, we introduce the log-normal distribution function~\cite{Klypin2017}
\begin{align}
    P_\mathrm{LN}(\rho_n) = \frac{1}{\rho_n\sqrt{2\pi\sigma^2_\mathrm{LN}}}\exp\left(-\frac{(\ln(\rho_n) + \sigma^2_\mathrm{LN}/2)}{2\sigma^2_\mathrm{LN}}\right)\,,
    \label{eq:fit_lognormal}
\end{align}
where $\sigma^2_\mathrm{LN}$ is the only free parameter. A fit for $\mathcal{N}=16.4$ is shown in the right panel of \cref{fig:dens_distribution},
however the log-normal distribution does not provide an accurate fit to the tails of the distribution function and does not align at all with the simulation results at later times. Consequently we use a power-law $P(\rho_n) \sim \rho_n^\alpha$ with slope parameter $\alpha$ truncated on small and large densities with exponential terms~\cite{Klypin2017} to model the distribution for $\rho_n\leq 0.2$ and another power-law with an exponential cutoff for $\rho_n > 0.2$. Specifically, the distribution functions are~\cite{Klypin2017}
\begin{align}
    P(\rho_n) = A\rho_n^\alpha\exp\left(-(\rho_1/\rho_n)^{1.1}\right) \exp\left(-(\rho_n/\rho_2)^{0.55}\right)
    \label{eq:fit_left}
\end{align}
for $\rho_n\leq 0.2$ and 
\begin{align}
    P(\rho_n) = B\rho_n^\beta\exp\left(-b\rho_n/\rho_3\right)
    \label{eq:fit_right}
\end{align}
for $\rho_n > 0.2$. In these two expressions $A$, $B$, $\alpha$, $\beta$, $b$, $\rho_1$, $\rho_2$ and $\rho_3$ are free fitting parameters. With slope parameters of $\alpha = 0.8$ and $\beta = -0.8$, the numerical data can be modeled accurately over the entire range of $\rho_n$. The combination of \cref{eq:fit_left,eq:fit_right} is also suited to describe the density distribution at earlier times, i.e. this approach is not limited to the $\mathcal{N}=20.0$ case.

\subsection{Halo Density Profiles}
\label{sec:density_profiles}

Dark matter halos developed via hierarchical structure formation are well-described by the Navarro-Frenk-White (NFW) profile~\cite{Navarro1996}
\begin{align}
    \rho_\mathrm{NFW}(r) = \frac{\rho_0}{r/r_s(1+r/r_s)^2}\,,
    \label{eq:nfw}
\end{align}
where $\rho_0$ is the characteristic density of a halo and $r_s$ the scale radius, and depend only weakly on the halo mass and cosmological parameters. The scale radius sets the size of the central region where $\rho \sim r^{-1}$ and $\rho\sim r^{-3}$ for $r \gg r_s$. The concentration of the halo is defined to be $c = r_\mathrm{vir}/r_s$.

We study the density profiles of the halos in our final snapshot in two  mass regimes. Since we are limited by spatial resolution, we focus on inflaton halos at the high-mass end of the HMF and separate between the mass regimes $M > 5\times 10^2\,M_{h,\mathrm{end}}$ and $M\in [2,3]\times 10^2\,M_{h,\mathrm{end}}$, at ${\cal{N}}=20$ in the $L = 50\,l_u$ simulation. 

\begin{figure}
    \centering
    \includegraphics{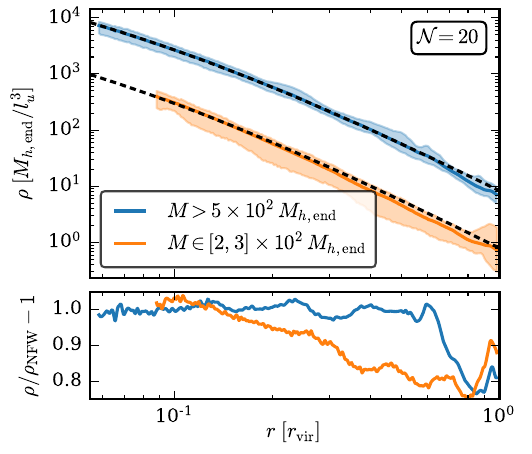}
    \caption{Averaged radial density profiles (upper panel) of 10 inflaton halos at ${\cal{N}}=20$ in two different mass bins (solid lines). The density profiles of the inflaton halos of masses $M\in [2,3]\times 10^2\,M_{h,\mathrm{end}}$ are rescaled by a factor of $10^{-1}$ for illustration purposes.  The shaded regions display the area filled by all 10 density profiles in the two mass bins. The black dashed lines represent NFW-fits, the deviations from the fits are shown in the lower panel. }
    \label{fig:densprofile}
\end{figure}

The averaged radial density profiles of ten inflaton halos in each mass sample are shown in \cref{fig:densprofile}.  Black dashed lines represent NFW-fits, given by \cref{eq:nfw} and the total range in the sample is shown by the coloured regions. Deviations from the NFW fits are displayed in the lower panel of \cref{fig:densprofile}. There is very good agreement for inflaton halos with $M>5\times 10^2\,M_{h,\mathrm{end}}$ deviating not more than $~\sim 20\%$ even for large $r$. For  lower-mass halos the NFW-fit is less accurate at large $r$ where the averaged density profile is slightly underdense compared to the NFW-fit. Nevertheless, the  profiles of the largest halos are NFW-like and exhibit concentrations $c\in [10,13.5]$. Likewise, consistent with standard dark matter simulations, the concentration parameter increases for a decreasing halo mass.   

\section{Conclusions and Discussion}
\label{sec:conclusions}

We have performed the largest-ever simulation of the smallest fraction of the Universe, evolving it from ${\cal{N}}=14$ to ${\cal{N}}=20$ $e$-folds after the end of inflation with a final physical box size of only $\sim 10^{-18}\,\mathrm{m}$, to explore the details of the gravitational fragmentation of the inflaton condensate during the early matter-dominated era following inflation.
We confirm that in the absence of prompt reheating small density fluctuations (as first analysed in detail in Ref~\cite{Musoke2019}) in the inflaton field collapse into gravitationally bound inflaton halos during this epoch.

Our results provide the first quantitative predictions for the mass and density statistics of the collapsed objects found in the very early universe during this phase. The inflaton halos have masses up to $\sim 10^3\,M_{h,\mathrm{end}}$ 
and their mass distribution is in agreement with the prediction of the mass-Peak Patch algorithm and with the Press-Schechter halo mass function as proposed in \cite{Niemeyer2019} after the first 1-3 $e$-folds of nonlinear growth.  As expected from dark matter N-body simulations, the density profiles of the inflaton halos are NFW-like with concentrations of $\mathcal{O}(10)$. Overall, the inflaton field reaches overdensities of close to $10^5$ in these simulations after three $e$-folds of nonlinear growth. 

The results do not depend on the precise form of the inflaton potential, provided (i) there is no strong resonant production of quanta in the immediate aftermath of inflation and (ii) couplings between the inflaton and other species leave space for a  long phase of matter-dominated expansion before thermalization.  

It is conceivable that dark matter is produced during the thermalization process  \cite{Chung:1998ua,Liddle:2006qz,Easther:2013nga,Fan:2013faa,Tenkanen:2016jic,Tenkanen:2016twd,Hooper:2018buz,Almeida:2018oid,Tenkanen:2019cik}. Even if this is not the case, the presence of gravitationally bound structures in the post-inflationary universe will modify the dynamics of reheating in ways that are yet to be properly explored.

The formation of gravitationally bound structures and their subsequent nonlinear evolution can source gravitational wave production~\cite{Jedamzik2010_grav}. Investigating a possible stochastic background sourced during this epoch is an obvious extension of this work, which will likely require N-body computations that extend further into the nonlinear epoch, potentially complemented by halo realizations of the mass-Peak Patch method. Thus, it will be possible to revisit the bounds obtained in \cite{Jedamzik2010_grav}. Likewise, the formation of solitons at the centres of the inflaton halos is currently unexplored, and will require full simulations (at least locally) of the corresponding Schr\"{o}dinger-Poisson dynamics. Moreover, given the duality between the underlying dynamics of the two eras any   supermassive black hole formation mechanism in the present epoch that is driven by dark-matter dynamics (and not baryonic physics) will have an early universe analogue with a mechanism for potential primordial black hole formation.

\section*{Acknowledgments}
We thank Christoph Behrens, Mateja Gosenca, Lillian Guo, Shaun Hotchkiss, Karsten Jedamzik, Emily Kendall, Nathan Musoke, and Bodo Schwabe for useful discussions and comments. We express a special thanks to Christoph Behrens for assistance with his \textsc{M3P} code. We  acknowledge  the \textsc{Yt} toolkit~\cite{yt} that  was used for our analysis of numerical data. JCN acknowledges funding by a Julius von Haast Fellowship Award provided by the New Zealand Ministry of Business, Innovation and Employment and administered by the Royal Society of New Zealand. RE acknowledges support from the Marsden Fund of the Royal Society of New Zealand.

\appendix

\section{Perturbation Equation and Power Spectrum}
\label{sec:perturbation_ps}

Decomposing the scalar field  into an inhomogeneous perturbation $\delta\varphi$ and a position-independent background $\overline\varphi$,
and working in spatially flat gauge, the perturbation equation in Fourier space is~\cite{Easther2010}
\begin{align}
    &\left(\frac{k^2}{a^2} + V''(\overline\varphi)  +2M_\mathrm{Pl}^{-2}\frac{\dot{\overline\varphi}}{H}V'(\overline\varphi) + M_\mathrm{Pl}^{-4}\frac{\dot{\overline\varphi}^2}{H^2}V(\overline\varphi) \right)\delta\varphi_k \nonumber\\
    &+ 3H\delta\dot\varphi_k + \delta\Ddot{\varphi}_k = 0\,.
    \label{eq:perturbation_eq}
\end{align}
To compute the power spectrum of density perturbations over a wide range of $k$ at the end of inflation, \cref{eq:perturbation_eq} has  to be solved numerically. However, it is possible to calculate the power spectrum at super- and subhorizon scales using the slow-roll and the Wentzel–Kramers–Brillouin (WKB) approximations, respectively. We work with the quadratic potential from \cref{eq:quadratic_potential}  in the spatially flat gauge where the curvature perturbation $\mathcal{R}$ is related to $\delta\varphi$ via $\mathcal{R} = -H\delta\varphi/\dot{\overline\varphi}$.

Modes that left the horizon during slow-roll inflation can be handled using the slow-roll approximation, i.e. $\varepsilon \ll 1$ and $\eta \ll 1$. In terms of the potential $V(\varphi)$, it is $\eta = M^2_\mathrm{Pl}  V''/V = m^2/(3H^2) \ll 1$ and thus the inflaton mass $m\ll H$ can be neglected. Since $\varepsilon = -\dot H/H^2 = \dot{\overline\varphi}^2/(2H^2M^2_\mathrm{Pl})\ll 1$ during slow-roll inflation, $H$ is constant and the $\dot{\overline\varphi}/H$ terms in \cref{eq:perturbation_eq} can be dropped. Hence, the perturbation equation reduces in this case to
\begin{align}
    \delta\Ddot{\varphi}_k + 3H\delta\dot\varphi_k + \frac{k^2}{a^2}\delta\varphi_k = 0\,.
\end{align}
One finds that a solution to this equation is given by 
\begin{align}
    \delta\varphi_k = \frac{1}{a\sqrt{2k}}\left(1 + i\frac{aH}{k}\right)\exp\left(\frac{ik}{aH}\right)\,,
\end{align}
and 
\begin{align}
    \mathopen|\delta\varphi_k\mathclose|^2 = \frac{1}{2ka^2}\left(1 + \frac{a^2H^2}{k^2}\right) = \frac{H^2}{2k^3}\left(1 + \frac{k^2}{a^2H^2}\right)\,, 
\end{align}
so $\delta\varphi_k\rightarrow H/(2k^3)^{1/2}$ for superhorizon scales $k\ll aH$.
The dimensionless curvature power spectrum at horizon crossing ($k=a_\ast H_\ast$) is
\begin{align}
    \Delta^2_\mathcal{R}(k) = \frac{k^3}{2\pi^2}\mathopen|\mathcal{R}_k\mathclose|^2 = \frac{H_\ast^2}{(2\pi)^2}\frac{H_\ast^2}{\dot{\overline\varphi}^2}\,,
\end{align}
and depends only weakly on scale $k$.

We now consider non-relativistic modes ($k/a \ll m$) that never leave the horizon ($k\gg aH$).
The slow-roll approximation is not applicable but 
the $V(\overline\varphi)\dot{\overline\varphi}^2/H^2$ term from  \cref{eq:perturbation_eq} can be omitted as it decays as $a^{-3}$ in the post-inflationary epoch while the $V'(\overline\varphi)\dot{\overline\varphi}/H$ term scales as $a^{-3/2}$.
Hence, the equation of motion for the subhorizon scales reads~\cite{Easther2010}
\begin{align}
    \delta\Ddot{\varphi}_k + 3H\delta\dot\varphi_k + \left(\frac{k^2}{a^2} + m^2 + 6Hm\sin(2mt)\right)\delta\varphi_k = 0\,.
    \label{eq:perturbation_eq_subhorizon}
\end{align}
In the absence of parametric resonance ($k/a \gg (3mH)^{1/2}$~\cite{Jedamzik:2010hq}) the last term in \cref{eq:perturbation_eq_subhorizon} can be neglected and one can use the leading-order WKB approximation to find
\begin{align}
    \delta\varphi_k = \frac{1}{a^{3/2}\sqrt{2\omega(t)}}\exp\left(i\int\omega(t)\,\mathrm{d}t\right)\,
\end{align}
with 
\begin{align}
    \omega(t) = \left(\frac{k^2}{a^2} + m^2\right)^{1/2}\,.
\end{align}
Since $k/a\ll m$, one can make the ansatz
\begin{align}
    \delta\varphi_k(t) = \frac{1}{a^{3/2}\sqrt{2m}}\left(A(t)e^{-imt} + B(t)e^{imt}\right)\,.
\end{align}
Plugging this ansatz into \cref{eq:perturbation_eq_subhorizon}, the perturbation equation can be written as a system of coupled equations for $A$ and $B$ which can be transformed into a second order differential equation. Solving this equation for $A$ gives the result for $B$ which leads to (see Ref.~\cite{Easther2010} for further details)
\begin{align}
    \delta\varphi_k = \frac{15}{\sqrt{2m}}i\frac{a^{9/2}m^3H^3}{k^6}\cos(mt)
\end{align}
for subhorizon scales at the end of inflation. From this one can obtain $\mathopen|\mathcal{R}_k\mathclose|^2$ and thus~\cite{Easther2010}
\begin{align}
    \Delta_\mathcal{R}^2(k) = \frac{75m^5}{8\pi^2}\frac{H^6}{M_\mathrm{Pl}^2}\frac{a^9}{k^9}\,.
\end{align}
Since $H^6a^9$ is constant during the post-inflationary epoch it can be evaluated at any time, most conveniently at the end of inflation. 
The power spectrum of the density perturbations 
\begin{align}
    \Delta_m^2(k) = \frac{4}{25}\left(\frac{k}{a_\mathrm{end}H_\mathrm{end}}\right)^4\Delta_\mathcal{R}^2(k)
\end{align}
at the end of inflation thus scales as $\Delta_m^2\sim k^{-5}$ for subhorizon modes.

\section{Unit System for the Simulations}
\label{sec:unitsystem}
 
We take the physical size of the horizon $\mathcal{N}=20$ $e$-folds after the end of inflation to define the comoving length unit $l_u = e^{20}H_\mathrm{end}^{-1} = 1.51\times 10^{-20}\,\mathrm{m}$, where $H_\mathrm{end} = m/\sqrt{6}$. 
Since inflaton halos are expected to have $\mathcal{O}(\mathrm{g})$ masses~\cite{Niemeyer2019}, we choose the mass unit as $m_u = 10^{-3}\,\mathrm{kg}$. Taking the gravitational constant in the new unit system as $G= 1\,l_u^3/(m_u t_u^2)$, our time unit is 
 \begin{align}
    t_u = \left(\frac{1}{6.67\times 10^{-11}} (l_u/\mathrm{m})^3\, \mathrm{kg}/m_u\right)^{1/2}\mathrm{s}\,,
\end{align}
i.e. $t_u = 7.23\times 10^{-24}\,\mathrm{s}$. The Hubble parameter at the end of the simulation is $H_{20} = e^{-30}H_\mathrm{end} = 6.49\,t_u^{-1}$. Using that $\rho_{20} = 3H_{20}^2/(8\pi G)$, the energy density at the end of the simulation in physical units is $\rho_{20} = 5.02\,m_u/l_u^3$. We normalize the final scale factor to unity such that the initial scale factor corresponds to $a_\mathrm{init}/a_{20} = e^{-6}$.

\section{Initial Conditions from \textsc{Music}}
\label{sec:ics_music}

As input \textsc{Music} requires a transfer function which is related to the power spectrum $P(k)$ via 
\begin{align}
    P(k) = \sigma_8 k^{n_s} T^2(k)\,,
\end{align}
where $n_s=0.961$ is the constant power spectrum spectral index after inflation and $\sigma_8$ is the normalization of the power spectrum. In the standard cosmology it is $\sigma_8 = 0.811$ but since we use another power spectrum $\sigma_8$ has to be computed from 
\begin{align}
    \sigma_8^2 = \frac{1}{2\pi^2}\int W(k)^2k^2P(k)\,\mathrm{d}k\,,
    \label{eq:sigma8}
\end{align}
where $W(k)$ is a top-hat filter function in Fourier space:
\begin{align}
    W(k) = \frac{3j_1(kR_8)}{kR_8}\,.
\end{align}
Here, $R_8$ denotes a top-hat filter of radius $R_8=8\,l_u$ and $j_1$ is the first order spherical Bessel function 
\begin{align}
    j_1(x) = \frac{\sin(x) - x\cos(x)}{x^2}\,.
    \label{eq:bessel}
\end{align}
Solving the integral in \cref{eq:sigma8} and including the growth factor by adding a factor of $(a_{20}/a_\mathrm{init})^2 = (e^6)^2$ in the integral in \cref{eq:sigma8} gives $\sigma_8 = 1.71$.

To verify the initial conditions setup we compare the power spectrum that was inserted in \textsc{Music} with the power spectrum calculated from the initial density field in \textsc{Nyx}; these are shown in \cref{fig:PS_initial}. The power spectra agree over a wide range of $k$ but for large $k$ there are deviations from the input power spectrum which are explained by two effects. The first  is the sharp cut-off that arises at different $k$ for different grid sizes, determined by the Nyquist frequency $k_\mathrm{Ny} = \pi N_\mathrm{cell}/L$, where $N_\mathrm{cell}$ denotes the grid size and $L=50\,l_u$ is the size of the simulation box. For $N_\mathrm{cell}=512$, $k_\mathrm{Ny} \simeq 32.2\,l_u^{-1}$ and for $N_\mathrm{cell}=128$, $k_\mathrm{Ny} \simeq 8.05\,l_u^{-1}$. However, there is also an artefact arising from an interpolation used in \textsc{Nyx} to compute the particle mass density. This process requires  roughly 3-4 cells and the deviations thus become apparent at $k \sim k_\mathrm{Ny}/3$. For larger $N_\mathrm{cell}$ the deviation from the input power spectrum kicks in at larger $k$, as seen in \cref{fig:PS_initial}.

\section{Spatial Resolution of N-Body Simulations}
\label{sec:resolution}

We need sufficient spatial resolution to accurately determine the  halo mass function and the density profiles of inflaton halos. Conversely, larger boxes are needed for longer runs but increasing  the box size in order to evolve the simulation for a longer time (and to get more massive inflaton halos) at a fixed grid size reduces the spatial resolution.

For our choice of $L=50\,l_u$ with $512^3$ particles we achieve a spatial resolution of $\Delta x = 9.8\times 10^{-2}\,l_u$, ensuring we can resolve the density profiles of the highest-mass halos. However, the NFW profile is unresolved for most of the inflaton halos in our volume; see \cref{sec:density_profiles}. As seen in \cref{fig:hmf}, the N-body HMF at $\mathcal{N}=19.6$ exhibits a turnover that is no longer fully resolved at $\mathcal{N}=20.0$. Consequently, we have saturated the limits on the spatial resolution for these computations.

\section{Identification of Spurious Halos from Artificial Fragmentation}
\label{sec:spurious_halos}

To identify spurious halos and to remove them from the halo catalogue, we follow the procedure from Ref.~\cite{Lovell2014}. In a first step, we trace all the particles that are in a parent halo at a certain $\mathcal{N}$ to their positions at the initial snapshot. This collocation of particles is the so-called protohalo. As suggested in Refs.~\cite{Behroozi2012,Allgood2005,Zemp2011}, the appropriate way of computing the ellipsis parameters of a (proto)halo is via the shape tensor $S$. Since all of the N-body particles in our simulation have the same mass, the shape tensor is 
\begin{align}
    S_{ij} = \frac{1}{N} \sum_k x_{k,i} x_{k,j}\,,
\end{align}
where $x_{k,i}$ denotes the $i$th component of the position of the $k$th particle relative to the center of mass of the protohalo. The sorted eigenvalues ($\lambda_a$, $\lambda_b$, $\lambda_c$) of $S$ are related to the ellipsis parameters ($c\leq b\leq a$) of the protohalo via $a^2/3$, $b^2/3$ and $c^2/3$, and the sphericity of the protohalo is defined as $s=c/a = (\lambda_c/\lambda_a)^{1/2}$.

\begin{figure} 
    \centering
    \includegraphics[width=0.95\columnwidth]{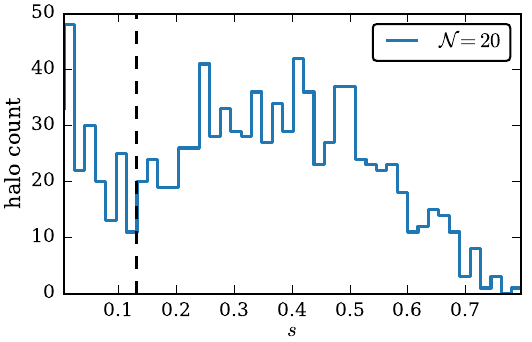}
    \caption{Histogram of the sphericity $s=c/a$ of all protohalos that belong to the identified halos ${\cal{N}}=20$ $e$-folds after the end of inflation. The dashed vertical black line marks the sphericity cut $s_\mathrm{cut} = 0.13$.}
    \label{fig:sphericity}
\end{figure}

We show the distribution of the sphericity $s$ of all protohalos corresponding to the identified halos at $\mathcal{N}=20$ in the $L=50\,l_u$ simulation in \cref{fig:sphericity}. Based on the strong increase of protohalos with $s<0.13$ we decided to set a cutoff $s_\mathrm{cut}=0.13$ below which protohalos are to be marked as spurious. The corresponding halos are removed from the halo catalogue.  This procedure removes most  but not all of the artificial halos. In principle, one should also filter out halos that do not have a counterpart in simulations of the same initial conditions but with a different spatial resolution. However, we only make the sphericity cut since the N-body HMF is additionally confirmed by the \textsc{M3P} results.

\section{Press-Schechter HMF with Sharp-$k$ Cutoff}
\label{sec:sharp_k}
\begin{figure}
    \centering
    \includegraphics[width=\columnwidth]{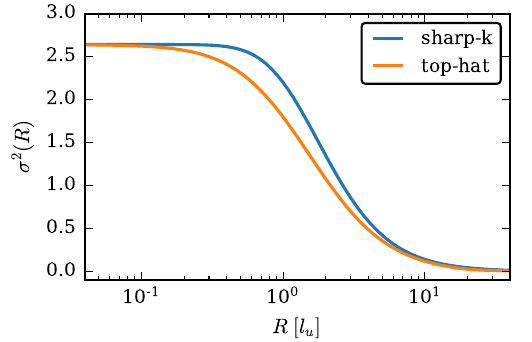}
    \caption{Variance of density perturbations $\sigma^2(R)$ using the sharp-$k$ and top-hat window function, respectively, see \cref{eq:W_sharp_k,eq:W_tophat}.}
    \label{fig:sigma2}
\end{figure}
As noted in \cref{sec:hmf}, we employed a sharp-$k$ filter to calculate the PS-HMF -- similar to the procedure used with WDM. This is a top-hat window function with a  radius $R$,  defined in Fourier space as 
\begin{align}
    W_k(kR) = \Theta(1-kR)\,.
    \label{eq:W_sharp_k}
\end{align}
This is in contrast to a top-hat window function in real space, with a filter scale $R_T$  given by
\begin{align}
    W_T(kR_T) = \frac{3j_1(kR_T)}{kR_T}
    \label{eq:W_tophat}
\end{align}
in Fourier space and $j_1$ is the first order spherical Bessel function from \cref{eq:bessel}.
Because of its sharp boundaries in real space, it is straightforward to define a mass $M_T = 4\pi\bar\rho R_T/3$ to the filter scale $R_T$.

However the sharp-$k$ filter has contributions on all scales in real space. This makes it difficult to assign a mass to the filter scale and a free parameter $\varepsilon$ is added to the mass assignment in \cref{eq:mass_sharpk}, chosen so that the PS-HMF matches the numerical simulations. Similarly to Refs.~\cite{Schneider2015,Benson2013,Kulkarni2020}  $\varepsilon=2.5$ provides a good fit to the data and does not vary with time. 

The critical density $\delta_c$ in \cref{eq:hmf_PS} must also be rescaled since $\delta_c$ was originally derived from simulations of spherical top-hat collapse. Following Ref.~\cite{Kulkarni2020}, we compared the variance of density perturbations on a filter scale $R$, 
\begin{align}
    \sigma^2(R) = \frac{1}{2\pi^2}\int k^2P(k)W^2(kR)\,\mathrm{d}k\,,
\end{align}
using the window functions from \cref{eq:W_sharp_k,eq:W_tophat}, see \cref{fig:sigma2}. We found that $\sigma^2(R)$ using the sharp-$k$ window function is larger by a factor of 1.44 for large $R$ (high masses). As discussed in Ref.~\cite{Kulkarni2020}, the critical density $\delta_c$ in \cref{eq:hmf_PS} has to be adjusted correspondingly. 

\bibliography{inflaton_halos_v2}

\end{document}